\documentclass{webofc}
\usepackage[varg]{txfonts}   
\begin{document}
\title{Present status of radiative and rare kaon decays}
%
%

\author{\firstname{Oscar} \lastname{Cat\`a}\inst{1}\fnsep\thanks{\email{oscar.cata@uni-siegen.de}} }

\institute{Center for Particle Physics Siegen (CPPS), Theoretische Physik 1, Universit\"at Siegen, Walter-Flex-Stra\ss e 3, D-57068 Siegen, Germany
          }

\abstract{%
I review recent developments in rare and radiative kaon decays from the theory side, with emphasis on those modes that are actively analyzed by the experimental collaborations. 
}
\maketitle
\section{Introduction}
\label{sec:1}

Rare and radiative kaon decays have a long history in particle physics as extremely useful tests of the Standard Model (SM) at long distances but also as probes of short-distance physics, including physics beyond the SM. 

Currently the main focus of attention is on the golden modes, $K^+\to \pi^+{\bar{\nu}}\nu$ and $K_L\to \pi^0{\bar{\nu}}\nu$, which are the object of devoted experiments by the NA62 and the KOTO collaborations, respectively. Such short-distance dominated decays can be computed in the SM using perturbation theory and are very clean probes of new physics. Aside from this, NA62 is regularly bringing up new analyses, either in the form of more precise measurements of long-distance dominated modes or as exclusion bounds for processes forbidden in the SM, such as lepton-flavor violating kaon decays. LHCb has also a kaon physics program that will soon start producing new results.

In this review I will concentrate on a number of rare and radiative kaon decays which have experienced recent developments both on the experimental and theoretical side. In most of the cases, these developments have greatly benefited from a very fluent communication between experimental and theoretical collaborations.

I will not report here on exotic or forbidden channels, which are a very active part of the NA62 program. For a review on these activities, I direct the reader to two recent talks at the EPS-HEP conference this year~\cite{Minucci:2021,Parkinson:2021}.  

\section{Current status of $K^+\to \pi^+{\bar{\nu}}\nu$ and $K_L\to \pi^0{\bar{\nu}}\nu$}
\label{sec:2}

The measurement of the golden modes $K^+\to \pi^+{\bar{\nu}}\nu$ and $K_L\to \pi^0{\bar{\nu}}\nu$ is presently the main experimental endeavour in experimental kaon physics. The NA62 experiment aims at a determination of $K^+\to \pi^+{\bar{\nu}}\nu$ within an accuracy of $5\%$ and the KOTO experiment is devoted to the first measurement of $K_L\to \pi^0{\bar{\nu}}\nu$. 

\begin{figure}[h]
\centering
\sidecaption
\includegraphics[width=5cm,clip]{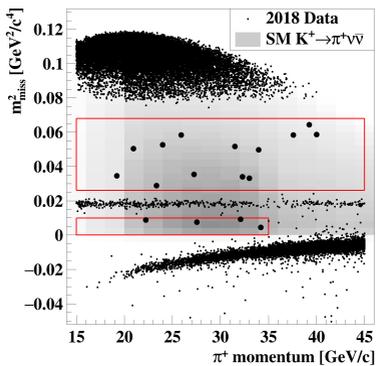}
\caption{Kinematical details of the NA62 data for $K^+\to \pi^+{\bar{\nu}}\nu$ corresponding to their 2018 dataset. Figure borrowed from~\cite{NA62:2021zjw}.}
\label{fig:1} 
\end{figure}

Both modes are theoretically extremely clean. Their prediction was revisited some years ago~\cite{Buras:2015qea}. $K^+\to \pi^+{\bar{\nu}}\nu$ is now computed including QCD and electroweak corrections at NLO for the top contribution, while for the charm contribution the QCD corrections are computed up to NNLO. The theoretical prediction yields
\begin{align}
{\cal{B}}(K^+\to \pi^+{\bar{\nu}}\nu)=(8.4\pm 1.0)\times 10^{-11}
\end{align}
This has to be compared with the most recent experimental determination from NA62 earlier this year, based on 20 signal events, mostly collected in 2018 (see fig.~\ref{fig:1}). Their final result is~\cite{NA62:2021zjw} 
\begin{align}
{\cal{B}}(K^+\to \pi^+{\bar{\nu}}\nu)=(10.6^{+4.0}_{-3.4}\pm 0.9)\times 10^{-11}\,,
\end{align}
which is the most precise determination to date. Their precision is foreseen to improve to $10\%$ with the data from Run 2, which started this year and will end in 2023. The final goal is to achieve a $5\%$ precision. From the results available right now one can conclude that sizeable deviations from the SM are to be excluded.

\begin{figure}[h]
\centering
\sidecaption
\includegraphics[width=5cm,clip]{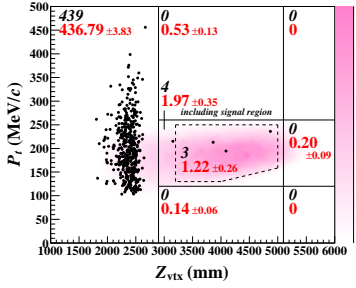}
\caption{Data from the KOTO experiment for the period 2016-2018. All the data points are consistent with the background estimates. Figure borrowed from~\cite{KOTO:2020prk}.}
\label{fig:2}  
\end{figure}

The theoretical prediction for $K_L\to \pi^0{\bar{\nu}}\nu$ was also updated in~\cite{Buras:2015qea}. The charm contribution is extremely suppressed and was neglected. The top contribution was computed including QCD and electroweak corrections at NLO. The final result reads
\begin{align}
{\cal{B}}(K_L\to \pi^0{\bar{\nu}}\nu)=(3.4\pm 0.6)\times 10^{-11}
\end{align}
This is an extremely challenging mode from the experimental side. The KOTO collaboration reported in 2018 preliminary results including 3 signal events from the data-taking period 2016-2018. After reexamination of the backgrounds these signals are now deemed consistent with backgrounds (see fig.~\ref{fig:2}). Their result from the analysis of the 2016-2018 data set is~\cite{KOTO:2020prk}
\begin{align}
{\cal{B}}(K_L\to \pi^0{\bar{\nu}}\nu)< 4.9\times 10^{-9}\,,
\end{align}
which is slightly higher than the bound reported for the 2015 data, ${\cal{B}}(K_L\to \pi^0{\bar{\nu}}\nu)< 3.0\times 10^{-9}$.

\section{$K\to \pi\gamma^*$ and $K\to \pi\pi\gamma^*$}
\label{sec:4}

Radiative 3- and 4-body decays are long-distance dominated modes, with the main contribution coming from Bremsstrahlung. The subleading hadronic contributions are typically termed electric and magnetic contributions, depending on their transformation properties under parity. These modes are important as tests of the Standard Model at low energies, which is described by Chiral Perturbation Theory (ChPT) in the $\Delta S=1$ sector:
\begin{align}
{\cal{L}}_{\Delta S=1}=G_8 f_{\pi}^4~{\mathrm{tr}}\big[\lambda_6 D_{\mu}U^{\dagger}D^{\mu}U\big]+G_8f_{\pi}^2\sum_j N_j\, W_j(U,D_{\mu}U,\lambda_6)+{\cal{O}}(p^6)\,,
\end{align}
where $U={\mathrm{exp}}\left[\frac{i}{f_{\pi}}\phi^a\tau^a\right]$ is the chiral matrix containing the Goldstone fields $\phi^a$. At NLO there are 37 unknown coefficients $N_j$, which capture the effects of nonperturbative hadronic states at the kaon mass scale. The general form of the amplitudes for these decay modes is therefore
\begin{align}
{\cal{M}}(K\to X\gamma^{(*)})=\underbrace{{\cal{M}}_B({\cal{O}}(p^2))}_{\mathrm{Brems.}}+ \underbrace{{\cal{M}}_E({\cal{O}}(p^4))}_{\mathrm{electric,\, CP-even}}+\underbrace{{\cal{M}}_M({\cal{O}}(p^4))}_{\mathrm{magnetic,\, CP-odd}}\,,
\end{align}  
where the electric and magnetic contributions are functions of the $N_j$ low-energy couplings. These couplings are related to the slopes of the differential decay rates and can be accessed by measuring the interference term between Bremsstrahlung and electric/magnetic emission. 

Experimentally this is a rather fertile field, with efforts that started roughly 15 years ago and are still ongoing. The present experimental situation can be summarized as follows:
\begin{equation}\label{eq-exp}
\begin{array}{lrlrlr}
K^{\pm}\to \pi^{\pm}\gamma^* & [10^{-7}]_{3\%}\,; & K_S\to \pi^0\gamma^* & [10^{-9}]_{50\%}\,; & K_L\to \pi^0\gamma^* & [< 10^{-10}] \\
\noalign{\vskip 2mm} 
K^{\pm}\to \pi^{\pm}\pi^0\gamma & [10^{-6}]_{7\%}\,; & K_S\to \pi^+\pi^-\gamma & [10^{-3}]_{3\%}\,; & K_L\to \pi^+\pi^-\gamma & [10^{-5}]_{4\%} \\ 
\noalign{\vskip 2mm}
K^{\pm}\to \pi^{\pm}\gamma\gamma & [10^{-6}]_{6\%}\,; & K_S\to \pi^0\gamma\gamma & [10^{-8}]_{37\%}\,; & K_L\to \pi^0\gamma\gamma & [10^{-6}]_{3\%} \\ 
\noalign{\vskip 2mm}
K^{\pm}\to \pi^{\pm}\pi^0\gamma^* & [10^{-6}]_{3\%}\,; & K_S\to \pi^+\pi^-\gamma^* & [10^{-5}]_{3\%}\,; & K_L\to \pi^+\pi^-\gamma^* & [10^{-7}]_{6\%}\,,
\end{array}
\end{equation}
where the value in brackets indicates the order of magnitude of each branching ratio and the percentage the current experimental precision.

In the following I will discuss two recent lines of research related to these decay modes. The first one describes a strategy, mostly using experimental input, to determine in a closed form the set of NLO coefficients relevant for radiative kaon decays. The second line of research is more ambitious and aims at a theoretical prediction of the NLO coefficients.

Before discussing these issues I would like to emphasize that radiative 3-body but especially 4-body decays can also be used as probes of short-distance physics, within and beyond the Standard Model. 4-body decays depend on 5 kinematic variables and therefore have very rich kinematical distributions. The differential amplitudes can be parametrized as
\begin{align}\label{angular}
\frac{d^5\Gamma}{dE_{\gamma}^*dT_c^*dq^2d\cos\theta_{\ell} d\phi}&={\cal{A}}_1+{\cal{A}}_2\sin^2\theta_{\ell}+{\cal{A}}_3\sin^2\theta_{\ell}\cos^2\phi+{\cal{A}}_4\sin2\theta_{\ell}\cos\phi\nonumber\\
&+{\cal{A}}_5 \sin\theta_{\ell}\cos\phi+{\cal{A}}_6 \cos\theta_{\ell}+{\cal{A}}_7\sin\theta_{\ell}\sin\phi\nonumber\\
&+{\cal{A}}_8\sin 2\theta_{\ell}\sin\phi+{\cal{A}}_9\sin^2\theta_{\ell}\sin 2\phi\,,
\end{align}
where ${\cal{A_j}}$ are form factors that depend on the dynamical variables $E_\gamma^*$, $T_c^*$ and $q^2$. The first line contains the terms responsible for $P$-conserving interactions, with only the first three terms contributing to the total cross section. The second and third lines instead describe $P$-violating interactions induced by short-distance and long-distance dynamics, respectively. They can be isolated with appropriate angular asymmetries and most of them yield very clean tests of new physics.

\subsection{Towards a determination of the radiative NLO chiral couplings}

Out of the 37 $N_j$ NLO operators of the $\Delta S=1$ ChPT Lagrangian, radiative kaon decays are sensitive to a rather small subset of them, namely $W_{14}-W_{18}$ (CP-even sector) and $W_{28}-W_{31}$ (CP-odd sector).

The observation made in~\cite{Cappiello:2017ilv} was that an individual determination of them all was feasible after the measurement of the electric interference terms either of $K^{\pm}\to \pi^{\pm}\pi^0\gamma^{*}$ or of $K_S\to \pi^+\pi^-\gamma^{*}$ (in combination with $K_L\to \pi^+\pi^-\gamma^{*}$). These decay modes were already in the programs of NA48/2 and LHCb and the observation thus provided an extra motivation for their analysis.

Out of the measured radiative kaon modes listed in~(\ref{eq-exp}) there are only 4 non-redundant combinations of NLO parameters. In particular, for the modes $K^\pm\to \pi ^{\pm} \gamma^*$, $K_{S}\to \pi ^{0} \gamma^*$, $K^{\pm }\to \pi ^{\pm }\pi ^{0}\gamma$ and $K^{\pm }\to \pi ^{\pm}\gamma\gamma$ one finds the combinations
\begin{align}
{\cal{N}}_E^{(1)}&\equiv N_{14}^r-N_{15}^r=\frac{3}{64\pi^2}\left(\frac{1}{3}-\frac{G_F}{G_8}a_+-\frac{1}{3}\log\frac{\mu^2}{m_K m_{\pi}}\right)-3L_9^r=-0.0167(13)\nonumber\\
{\cal{N}}_S&\equiv 2N_{14}^r+N_{15}^r=\frac{3}{32\pi^2}\left(\frac{1}{3}+\frac{G_F}{G_8}a_S-\frac{1}{3}\log\frac{\mu^2}{m_K^2}\right)=+0.016(4)\nonumber\\
{\cal{N}}_E^{(0)}&\equiv N_{14}^r-N_{15}^r-N_{16}^r-N_{17}=-\frac{|{\cal{M}}_K| f_{\pi}}{2G_8}X_E=+0.0022(7)\nonumber\\
{\cal{N}}_0&\equiv N_{14}^r-N_{15}^r-2N_{18}^r=\frac{3}{128\pi^2}{\hat{c}}-3(L_9^r+L_{10}^r)=-0.0017(32)\,,
\end{align} 
where in the second equality on each line one can see how the NLO counterterms are related to the slopes $a_+$, $a_S$, $X_E$ and ${\hat{c}}$. The numbers on the right-hand side are obtained by using the experimental numbers for the slopes.

\begin{figure}[h]
\centering
\sidecaption
\includegraphics[width=5cm,clip]{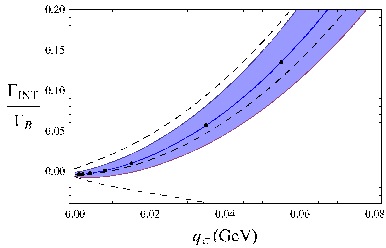}
\caption{Ratio between the interference and the Bremsstrahlung contributions for the $K^{\pm}\to \pi^{\pm}\pi^0 e^+e^-$ decay. The solid line and the error bar correspond to the values displayed in the main text. The upper and lower dashed curves correspond to setting ${\cal{N}}_E^{(0)}=0$ and ${\cal{N}}_E^{(0)}=0$, respectively. Figure borrowed from~\cite{Cappiello:2017ilv}.}
\label{fig:3}    
\end{figure}

One can easily see that $N_{14}$, $N_{15}$ and $N_{18}$ can be individually determined, but only the combination $N_{16}+N_{17}$ is constrained. When $K^{\pm}\to \pi^{\pm}\pi^0 e^+e^-$ is considered, one is sensitive to ${\cal{N}}_E^{(0)}$, ${\cal{N}}_E^{(1)}$ and to a new independent combination of counterterms, namely
\begin{align}
{\cal{N}}_E^{(2)}&=N_{14}^r+2N_{15}^r-3(N_{16}^r-N_{17})
\end{align}
This combination leads to ${\cal{N}}_E^{(2)}=+0.089(11)+6N_{17}$ when only polynomial terms are considered. Since $N_{17}$ is predicted to be small by different hadronic models, one can assume that indeed ${\cal{N}}_E^{(2)}\sim +{\cal{O}}(10^{-1})$ and it is sizeable. The interference term has a characteristic pattern than depends mostly on ${\cal{N}}_E^{(0)}$ and ${\cal{N}}_E^{(2)}$. This is illustrated in fig.~\ref{fig:3}, where the upper and lower dotted lines correspond to setting ${\cal{N}}_E^{(0)}=0$ and ${\cal{N}}_E^{(2)}=0$, respectively.   

Even though $K^{\pm}\to \pi^{\pm}\pi^0 e^+e^-$ is a 4-body decay and has the corresponding phase-space suppression, the decay rate is comparable in size to $K^+\to \pi^+\gamma^{*}$. The challenge is to extract information on the structure-dependent terms. The magnetic decay width is 70 times smaller than the Bremsstrahlung, and for the interference term one typically needs to have a sensitivity below the percent level. However, the Bremsstrahlung is mostly localized at very small values of the dilepton invariant mass. Using cuts on this variable one can then increase the required sensitivity above the percent level.

An analysis of this decay has been published recently~\cite{NA482:2018gxf}, based on the complete data sets of the NA48/2 experiment. The Bremsstrahlung and magnetic contributions could be measured and agreement was found with the theoretical prediction. Unfortunately, in order to pin down the interference term more statistics would be needed. Progress on this side will have to come from a new experiment.

An important alternative to determine the set of NLO counterterms is to combine data on $K_{S}\to \pi^+\pi^- e^+e^-$ and $K_{L}\to \pi^+\pi^- e^+e^-$. Information on the decay rates exists for both modes, but not for the interference term, which would require typically a percent precision. Right now a new measurement of $K_{S}\to \pi^+\pi^- e^+e^-$ is in the LHCb agenda but I am not aware of any new measurement planned for $K_{L}\to \pi^+\pi^- e^+e^-$. One should probably stress that the counterterm structure for $K_S\to \pi^+\pi^-e^+e^-$ can be predicted from other radiative kaon decays to be 
\begin{align}
BR(K_S\to \pi^+\pi^-e^+e^-)=\underbrace{4.74\cdot 10^{-5}}_{\text{Brems.}}+\underbrace{4.39\cdot 10^{-8}}_{\text{Int.}}+\underbrace{1.33\cdot 10^{-10}}_{\text{DE}}\,,
\end{align}
but in order to determine $N_{16}$ and $N_{17}$ one needs to measure the slope of $K_{L}\to \pi^+\pi^- e^+e^-$.

\subsection{$K^+\to\pi^+\ell^+\ell^-$}

In the previous section the NLO counterterms were extracted from experiment. However, they can in principle be estimated theoretically. In~\cite{DAmbrosio:2018ytt} a new method was proposed and applied to $K^+\to\pi^+\ell^+\ell^-$. 

\begin{figure}[h]
\centering
\includegraphics[width=5.5cm,clip]{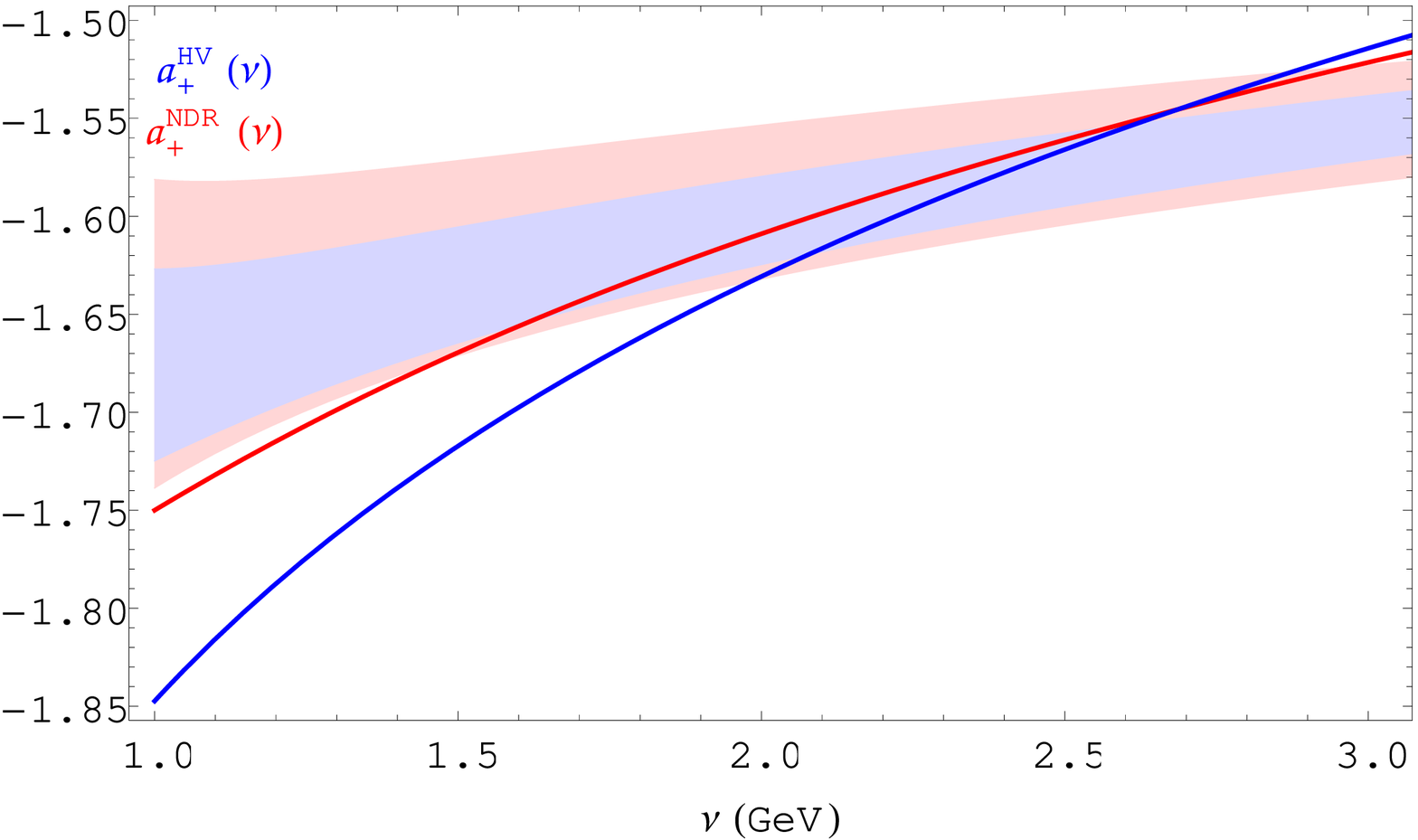}
\hskip 0.2cm
\includegraphics[width=5.5cm,clip]{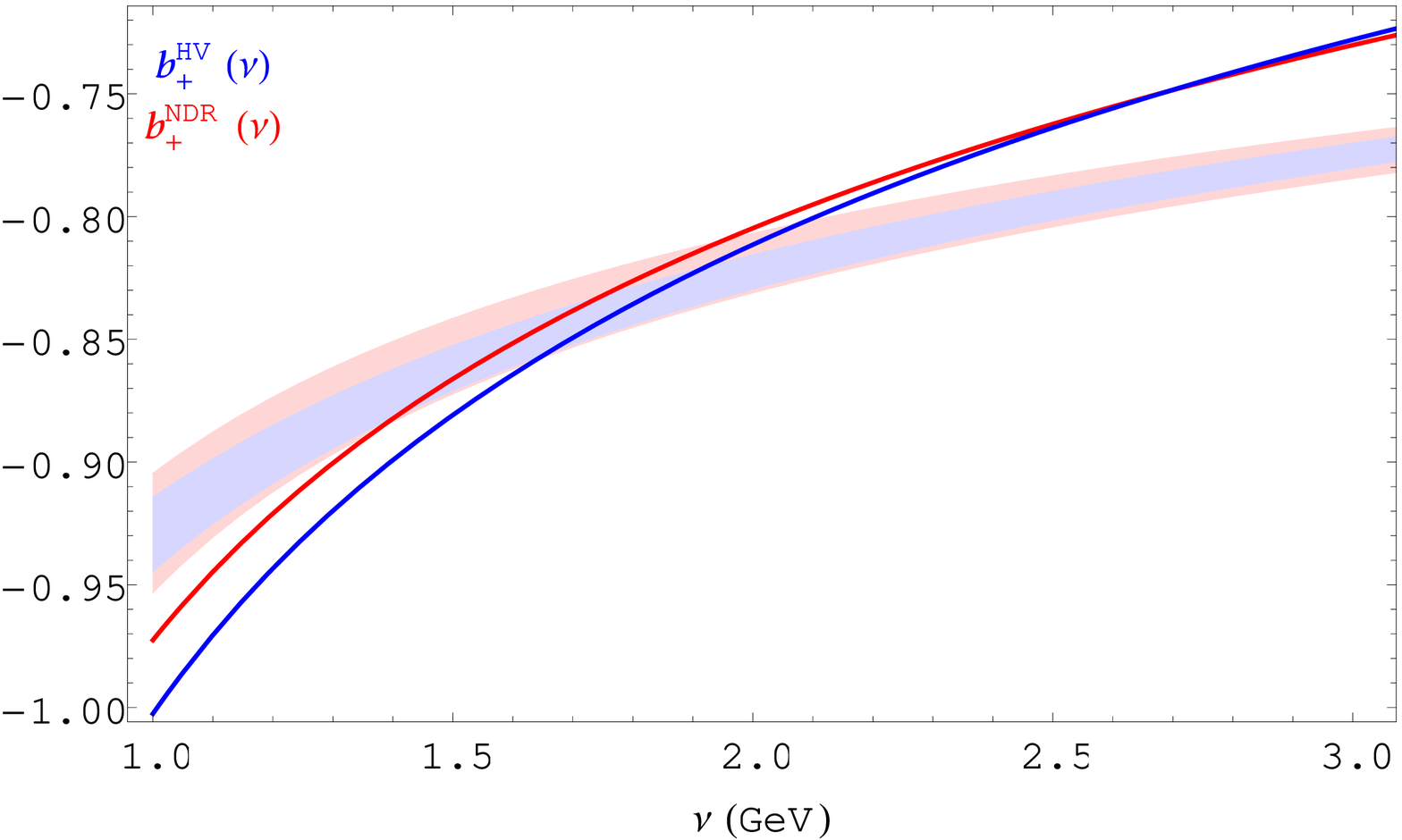}
\caption{Scale dependence of the slope parameters $a_+$ and $b_+$. Figure borrowed from~\cite{DAmbrosio:2018ytt}.}
\label{fig:4}    
\end{figure}

It is customary to parametrize this differential decay rate as
\begin{align}
\frac{d^2\Gamma}{dzdx}=\frac{\alpha^2 m_K}{(4\pi)^5)}f(x,z)|W(z)|^2, \qquad W(z)=G_Fm_K^2(a_++b_+z)+W^{\pi\pi}(z)\,,
\end{align}
where $a_+$ and $b_+$ are slope parameters. A determination of such parameters requires a description of $W(z)$ for all values of $z$. As a first approximation, the authors of ~\cite{DAmbrosio:2018ytt} suggested the ansatz
\begin{align}
W(z)=W^{\pi\pi}(z)+W^{\rm{res}}(z;\nu)+W^{\rm{SD}}(z;\nu)\,,
\end{align}
where the first term is the pion rescattering contribution, which can be extracted from experimental data, and the last piece is the short-distance tail, which can be computed with the operator product expansion. The intermediate-energy region is modelled with an infinite tower of states, whose parameters are chosen to match the high- and low-energy ends. If this matching is done at 1-loop level, one is sensitive to the scale dependence of the slope parameters. This scale-dependence is illustrated in fig.~\ref{fig:4}.

The value found for $a_+$ with this method is clearly off, but one should take into account that the model was rather crude. In particular, no kaon rescattering effects were taken into account. Work in this direction is currently underway. 

The $K^+\to\pi^+\mu^+\mu^-$ mode has been recently reanalyzed by NA62 with the result~\cite{Bician:2020ukv}
\begin{align}
a_+=-0.592(13)(7)(1);\qquad b_+=-0.699(46)(35)(3)
\end{align}
The results are still preliminary, but show a preference for negative values of the slopes. There exists a positive solution at $a_+=0.368$, $b_+=2.045$, but it is statistically disfavored. The existence of two sets of solutions with similar statistical significance is a feature of the data which was already pointed out in~\cite{DAmbrosio:2018ytt}. The new result of NA62 agrees with previous determinations and, once compared with the existing data for the electron channel, shows no indication of lepton-flavor universality violation. 

\begin{figure}[h]
\centering
\includegraphics[width=5cm,clip]{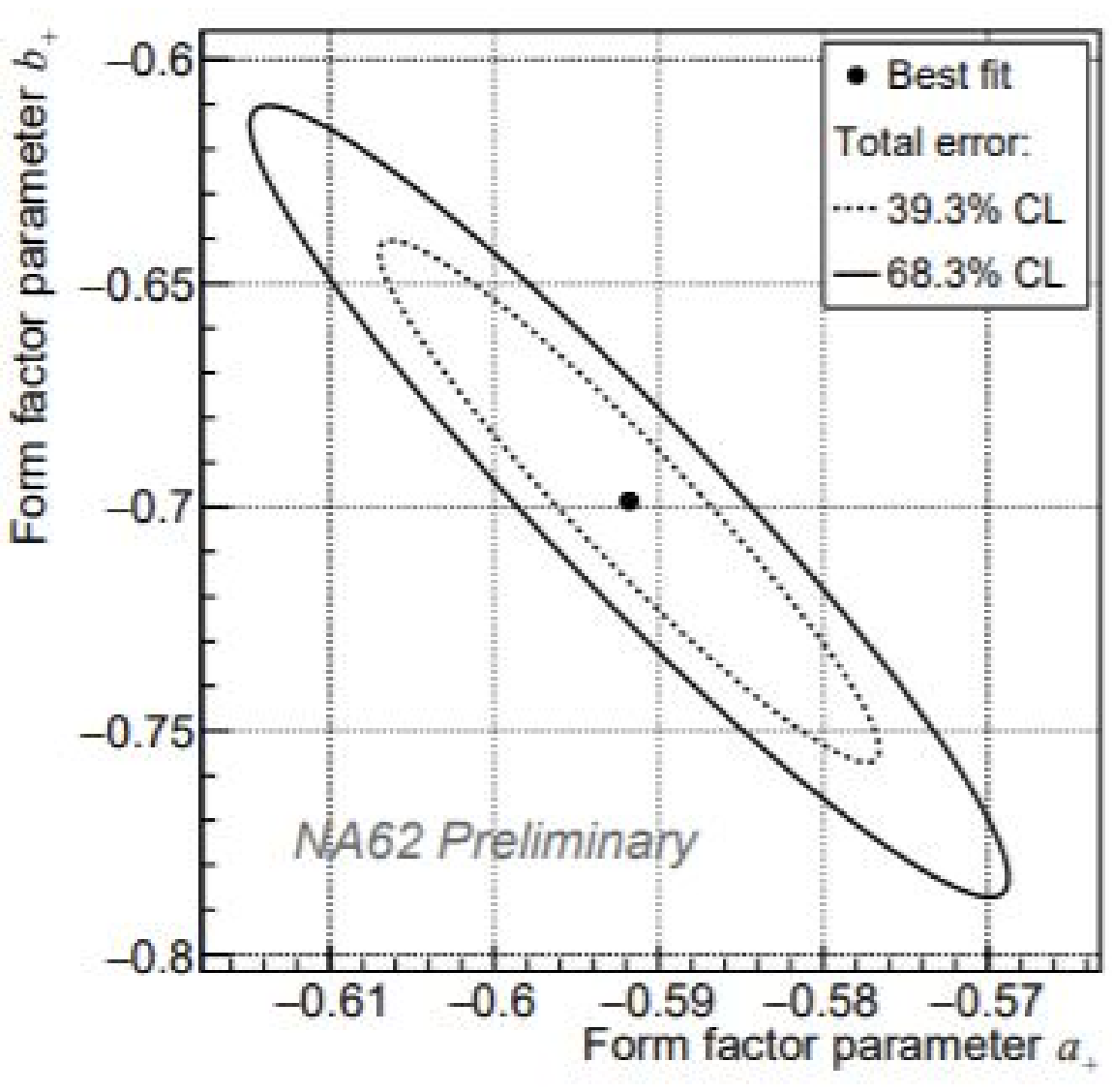}
\hskip 0.8cm
\includegraphics[width=4.7cm,clip]{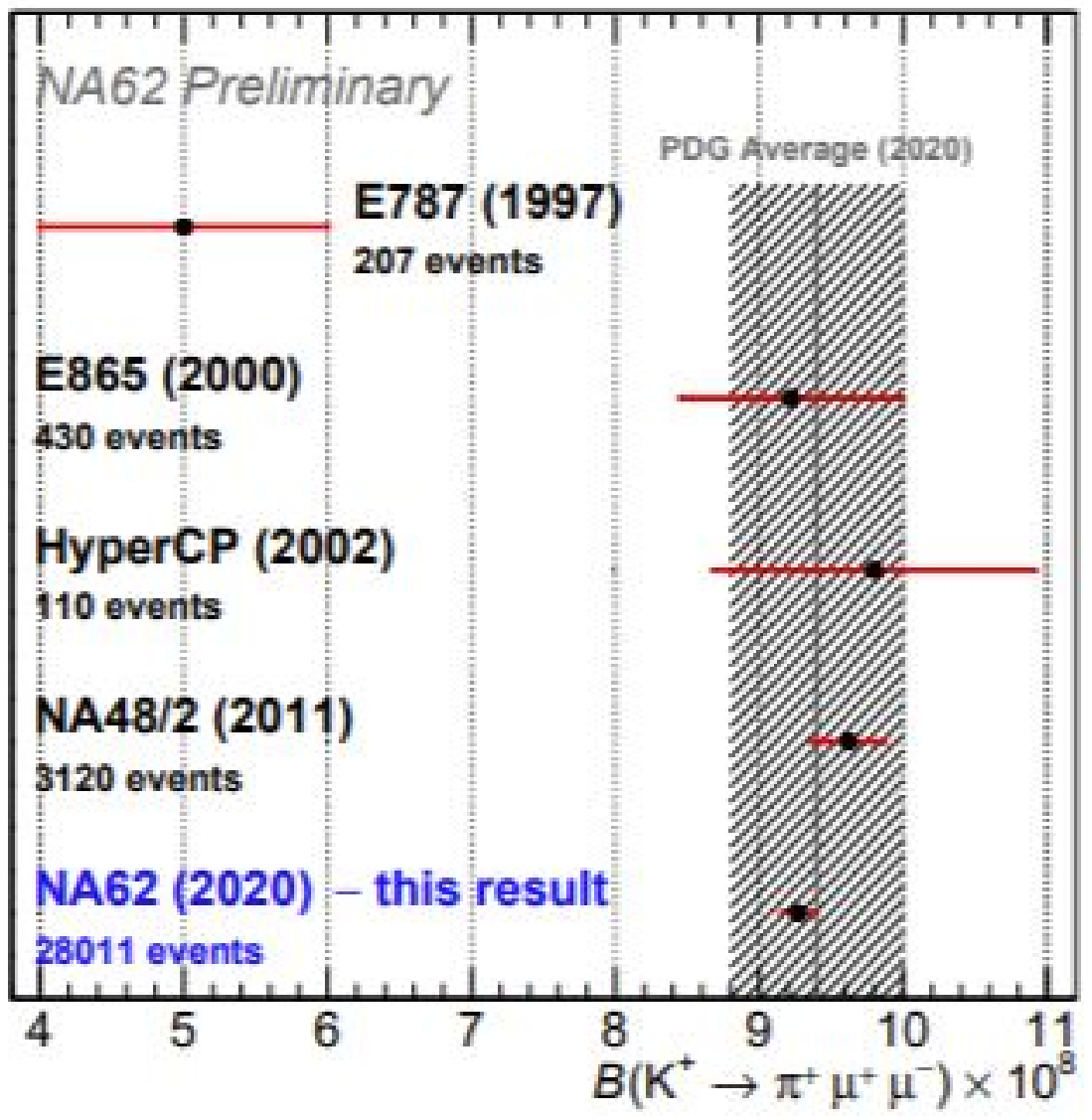}
\caption{Left panel: NA62 results for the slope parameters $a_+$ and $b_+$. Right panel: NA62 new determination of the branching ratio for $K^+\to\pi^+\mu^+\mu^-$ and comparison with previous results. Figures borrowed from~\cite{Bician:2020ukv}.}
\label{fig:5}       
\end{figure}

\section{$K^+\to \pi^0 e^+\nu\gamma$}
\label{sec:3}

I will finally review briefly the status of the $K_{\ell 3\gamma}$ decays. These modes are interesting for different reasons. Rather generically, they can be used as tests of the Standard Model at low energies, but the charged modes are also sensitive to the chiral anomaly and can probe T violation. 

Similarly to $K\to\pi\gamma^{(*)}$ and $K\to\pi\pi\gamma^{(*)}$, these modes are long-distance dominated and can be analyzed in the context of ChPT. The Bremsstrahlung piece is the dominant contribution (of ${\cal{O}}(p^2)$ in the chiral counting) and through Low's theorem it can be expressed in terms of the $K_{\ell 3}$ decay modes. The structure-dependent pieces start at ${\cal{O}}(p^4)$ in the chiral expansion and are typically probed with the ratios  
\begin{align}
R_j=\frac{{\cal{B}}(K\ell3\gamma_j)}{{\cal{B}}(K\ell3)}\,,
\end{align}
where the subindex $j$ denotes some specific cuts on the photon energy and the angle spanned between the lepton and the photon. It is common to use the benchmark ratios $R_1(E_\gamma>10$ MeV; $\theta>10^o)$, $R_2(E_\gamma>30$ MeV; $\theta>20^o)$ and $R_3((E_\gamma>10$ MeV; $0.6<\cos\theta<0.9)$.

The decay mode $K_{e3\gamma}$ has been recently measured by the OKA collaboration~\cite{OKA:2020zrs} and NA62~\cite{Brizioli:2021}. The results of the two collaborations for the three reference ratios $R_j$ can be found in Table~\ref{tab:1}. The values are compared with the theoretical estimates, which were carried out up to ${\cal{O}}(p^6)$ in ChPT~\cite{Kubis:2006nh}.

The results for $R_2$ and $R_3$ are largely compatible between NA62 and OKA, and show a notable deficit with respect to the theoretical result. The situation with $R_1$ is more puzzling: OKA shows a marked excess and NA62 a marked deficit with respect to the theoretical number. 

\begin{table}[h]
\centering
\caption{Experimental numbers for $R_j$ for the $K_{e3\gamma}$ decay compared to the theoretical estimate.}
\label{tab:1}       
\begin{tabular}{llll}
\hline
&\mbox{${\cal{O}}(p^6)$ ChPT~\cite{Kubis:2006nh}}&\mbox{OKA collaboration~\cite{OKA:2020zrs}}&\mbox{NA62 (preliminary)~\cite{Brizioli:2021}}  \\\hline
$R_1(\times 10^2)$&$1.804(21)$ &$1.990(17)(21)$ &$1.684(5)(10)$\\
$R_2(\times 10^2)$&$0.640(8)$ &$0.587(10)(15)$ &$0.599(3)(5)$\\
$R_3(\times 10^2)$&$0.559(6)$ &$0.532(10)(12)$ &$0.523(3)(3)$\\\hline
\end{tabular}
\end{table}
It is however hard to assess how serious these discrepancies are. The NA62 results are preliminary and one should wait for the published version of their analysis. The values released so far corresponds to the data sets of 2017 and 2018, so there is room to enhance their statistics. If their numbers are confirmed, one could claim a sizeable deficit for all ratios with respect to the ChPT estimate. The ${\cal{O}}(p^6)$ analysis was performed taking into account only the polynomial pieces and the NNLO low-energy constants were estimated with power-counting arguments. A discrepancy with experiment would motivate a revision of the theory estimate to include the non-polynomial pieces. A better estimate of the NNLO constants would be hard to achieve, given how poorly-known they are.

Right now, the $R_j$ ratios are the only realistic tests for this decay mode. Experiments are not sensitive to the T-odd parameter: the present upper bounds are two orders of magnitude larger than the theoretical prediction.

\section{Summary and outlook}
\label{sec:5}

Rare and radiative kaon decays are an active field of research thanks to the ongoing scientific programs of the NA62 and KOTO experiments. Most of the energies are currently devoted to the $K^+\to \pi^+{\bar{\nu}}\nu$ and $K_L\to \pi^0{\bar{\nu}}\nu$ modes. For the former, NA62 will soon achieve a $10\%$ precision, but sizeable new-physics effects are already excluded. For the latter, there exist so far upper bounds. In parallel, there is an intense experimental activity on exotic and forbidden modes by NA62. Regarding radiative kaon decays, there is still room for improvement on the experimental side, especially with the $K\to\pi\pi\gamma^{*}$ modes, where there is very little information aside from the branching ratios. The recent analysis of $K^+\to\pi^+\pi^0 e^+e^-$ has shown that the statistics collected by NA48/2 were not enough for a determination of the electric NLO chiral counterterms. LHCb will soon measure $K_S\to\pi^+\pi^-e^+e^-$, but it would be interesting to have also a more precise measurement of $K_L\to\pi^+\pi^-e^+e^-$ somewhere in the near future.   

\section*{Acnowledgements}
I would like to thank Evgueni Goudzovski and Rainer Wanke for kindly guiding me to the latest NA62 results.

\end{document}